# 1D photonic crystal direct bandgap GeSn-on-insulator laser

Hyo-Jun Joo,[1,a)] Youngmin Kim,[1,a)] Daniel Burt,[1] Yongduck Jung,[1] Lin Zhang,[1] Melvina Chen,[1] Samuel Jior Parluhutan,[1] Dong-Ho Kang,[1] Chulwon Lee,[2] Simone Assali,[3] Zoran Ikonic,[4] Oussama Moutanabbir,[3] Yong-Hoon Cho,[2] Chuan Seng Tan,[1] and Donguk Nam[1,b)]

[1] School of Electrical and Electronic Engineering, Nanyang Technological University, 50 Nanyang Avenue, Singapore 639798, Singapore

[2] Department of Physics and KI for the NanoCentury, Korea Advanced Institute of Science and Technology (KAIST), Daejeon 34141, Republic of Korea

[3] Department of Engineering Physics, École Polytechnique de Montréal, Montréal, C.P. 6079, Succ. Centre-Ville, Montréal, Québec H3C 3A7, Canada

[4] School of Electronic and Electrical Engineering, University of Leeds, Leeds LS2 9JT, UK

GeSn alloys have been regarded as a potential lasing material for a complementary metal-oxide-semiconductor (CMOS)-compatible light source. Despite their remarkable progress, all GeSn lasers reported to date have large device footprints and active areas, which prevent the realization of densely integrated on-chip lasers operating at low power consumption. Here, we present a 1D photonic crystal (PC) nanobeam with a very small device footprint of 7 μm$^2$ and a compact active area of ~1.2 μm$^2$ on a high-quality GeSn-on-insulator (GeSnOI) substrate. We also report that the improved directness in our strain-free nanobeam lasers leads to a lower threshold density and a higher operating temperature compared to the compressive strained counterparts. The threshold density of the strain-free nanobeam laser is ~18.2 kW cm$^{-2}$ at 4 K, which is significantly lower than that of the unreleased nanobeam laser (~38.4 kW cm$^{-2}$ at 4 K). Lasing in the strain-free nanobeam device persists up to 90 K, whereas the unreleased nanobeam shows a quenching of the lasing at a temperature of 70 K. Our demonstration offers an avenue towards developing practical group-IV light sources with high-density integration and low power consumption.

A practical group-IV light source is an indispensable component that has long been pursued to realize fully functional photonic-integrated circuits (PICs).[1,2] Since the first demonstration of Ge-based lasers,[3] there have been intensive research activities to realize Ge-based lasers by employing various methods including heavy n-type doping[4,5] and strain engineering.[6–11] More recently, GeSn alloys have shown great promise as potential complementary metal-oxide-semiconductor (CMOS)-compatible lasing materials because of their direct bandgap achieved at a high enough Sn content.[12–16] The first demonstration of optically pumped lasing in GeSn from a Fabry–Pérot type waveguide showed a lasing threshold density of 325 kW cm$^{-2}$ at 20 K.[17] However, the threshold density of this laser was too high for most practical applications. Since then, relentless efforts

---

a) Hyo-Jun Joo and Youngmin Kim contributed equally to this work.

b) Author to whom correspondence should be addressed: dnam@ntu.edu.sg

have been made to increase the Sn content for improving GeSn laser performance.[16,18–24] Recently, an operating temperature of 270 K in 20 at% Sn content has been reported by Zhou *et al*.[21] Although the operation is approaching room temperature, the high lasing threshold density of 364 kW cm$^{-2}$ of this GeSn laser still limits its practicality. More recently, Elbaz *et al*. presented a low threshold density of 0.8 kW cm$^{-2}$ at 25 K in a GeSn microdisk laser by combining tensile strain and the material quality advantages of low-Sn-content alloys.[23] Several studies have also proposed various approaches to improve the lasing threshold and operating temperature of GeSn lasers such as multi-quantum wells[25–28] and defect management.[29] Despite these remarkable achievements, the large active areas and device footprints of the aforementioned lasers require high power consumption and also limits a dense integration of lasers on a chip, which is an obstacle for the realization of energy-efficient light sources densely integrated on PICs.[30]

Thanks to their capability to achieve small mode volumes ($V_m$) and high quality ($Q$-) factors, photonic-crystal (PC) nanocavities have been regarded as a promising platform for realizing compact and energy-efficient laser sources.[30–34] While a two-dimensional (2D) PC laser has recently been reported,[35] the large area of the air hole array in the cavity (~900 μm$^2$) still restricts the realization of high-density integration. On the other hand, one-dimensional (1D) PC lasers may fulfil all the above-mentioned requirements for on-chip laser sources, such as small footprints and low power consumption.[36–40] Despite these remarkable advantages, there has not been any demonstration of a 1D PC laser in GeSn.

In this work, we demonstrate a GeSn-based 1D PC nanobeam laser with a very small device footprint of 7 μm$^2$. By using a high-quality GeSn-on-insulator (GeSnOI) substrate that allows releasing the limiting compressive strain thus improving the threshold and operating temperature in our nanobeam. Pump-power-dependent photoluminescence (PL) studies show a threshold density of 18.2 kW cm$^{-2}$ at 4 K for the released strain-free GeSn nanobeam, which is ~2 times lower than that of the unreleased GeSn nanobeam with compressive strain. The improved bandgap directness in the released GeSn nanobeam also enabled lasing action at higher operating temperatures up to 90 K compared to the unreleased laser device (<70 K). Our achievements pave the way toward practical laser sources with dense integration and extremely low power consumption for CMOS-compatible PICs.

Figure 1(a) presents a cross-sectional transmission electron microscopy (TEM) image of the fabricated GeSnOI substrate. An epitaxially grown GeSn layer with a Sn content of 10.6 at% was grown on a Si substrate with a Ge layer as a buffer using low-pressure chemical vapor deposition (LPCVD). Direct wafer bonding was performed to produce the GeSnOI substrate. A chemical-mechanical polishing (CMP) was used to remove the top defective layer, which was originally the interface between Ge and GeSn. After the entire process was completed, a high-quality GeSn layer with a thickness of 550 nm was obtained on a substrate comprising of Al$_2$O$_3$, SiO$_2$, and Si. The bonding process used in this study was optimized at a temperature lower than 225°C to avoid Sn segregation in the GeSn layer.

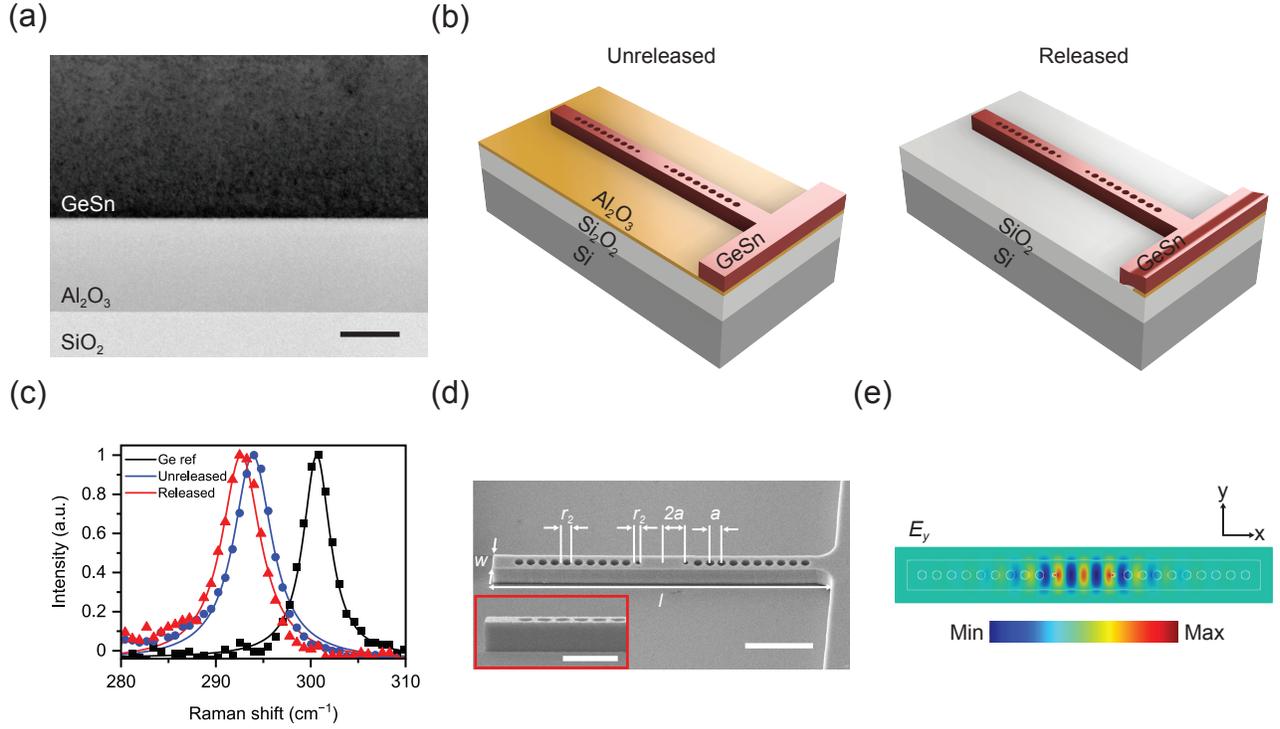

FIG. 1. (a) Cross-sectional TEM image of the GeSnOI substrate. Scale bar, 200 nm. (b) Schematics of the unreleased (left) and released (right) nanobeams. (c) Raman spectra of the unreleased and released nanobeams (bulk Ge for reference). (d) Tilted view SEM image of the released nanobeam. Scale bar, 2 μm. The magnified SEM image taken at a different angle is shown in the inset. Scale bar, 1 μm. (e) Electric field profile ($E_y$ component) of the nanobeam, simulated by using the FDTD method.

To see the effect of the released compressive strain on the lasing characteristics, we investigated unreleased and released 1D PC nanobeams as schematically shown in Fig. 1(b). The PC nanobeam structure was patterned by electron-beam lithography (EBL), followed by Cl$_2$ dry etch with reactive ion etching (RIE) to transfer the pattern to the GeSn layer for the unreleased nanobeams (left panel, Fig. 1(b)). To fabricate the released nanobeams (right panel, Fig. 1(b)), the Al$_2$O$_3$ layer underneath the GeSn layer was selectively wet-etched with 30 wt% potassium hydroxide (KOH) at 80°C. The sample was subsequently rinsed in deionized water and dried, which allowed the released nanobeam to be brought into contact with the SiO$_2$ layer owing to the capillary forces. The limiting compressive strain in GeSn can be significantly relaxed during the undercut process while the contact with the SiO$_2$ layer allows for adequate thermal management and strong optical confinement simultaneously.[41] To confirm the relaxation of compressive strain in the released nanobeam, Raman spectroscopy was conducted. For Raman spectroscopy measurements, a 532-nm laser was focused on the nanobeam using ×100 objective lens. The laser power was kept low (160 μW) to minimize any heating effect, which is evidenced in the pump-power-dependent measurements (Fig. S1, supplementary material). Figure 1(c) shows the experimentally measured Raman spectra of the fabricated devices at the center of the nanobeam. The Raman spectrum of a bulk Ge substrate is shown for reference. We observed a Raman peak shift of ~1.2 cm$^{-1}$ between the unreleased (~293.8 cm$^{-1}$) and released nanobeams (~292.6 cm$^{-1}$) due to the strain relaxation. The corresponding intrinsic compressive strain is calculated to be −0.23% by using a Raman-strain shift coefficient of 521 cm$^{-1}$.[42] Figure 1(d) presents a scanning electron microscopy (SEM) image of the released 1D PC nanobeam. The width ($w$) and length ($l$) of the nanobeam are 700 nm and 10 μm, respectively, making the device footprint as small as 7 μm$^2$. The PC cavity contains periodic air holes along the nanobeam with a fixed lattice constant ($a$) of 350 nm and three missing air holes at the center of the beam. To reduce the scattering losses in the cavity by providing a gradual refractive index variation, the radius of the inner ($r_1$) and outer ($r_2$) air holes are designed to be 0.16$a$ and 0.31$a$, respectively. The electric field profile ($E_y$) of the mode was simulated by the finite-difference time-domain (FDTD) method, as shown in Fig. 1(e). A strong optical confinement in the PC cavity is observed, and the calculated $Q$-factor and $V_m$ are 14600 and 0.76($\lambda/n$)$^3$, respectively. The resonance wavelength and refractive index in the simulation are 2194.08 nm and 4.25, respectively.

We performed PL measurements to analyze and compare the lasing characteristics of the unreleased and released nanobeams. The samples were mounted into a helium cryostat which operates at a temperature range between 4 K and 300 K. The two nanobeams were then optically pumped by a pulsed laser with a wavelength of 1550-nm, a repetition rate of 1 MHz, and a pulse width of 5 ns. A ×15 reflective objective lens was utilized to produce a laser spot size of ~13 μm and to collect the PL signal. The collected PL signal was guided to a Fourier transform infrared (FTIR) spectrometer with a spectral resolution of ~0.15 nm, and detected by an InGaAs detector.

Figures 2(a) and (b) present power-dependent PL spectra of unreleased and released nanobeams, respectively. The measurement was performed at a temperature of 4 K. The lasing modes at ~2135 nm and ~2204 nm are observed for unreleased and released nanobeams, respectively. We confirm that the emission from both nanobeams is attributed to the direct bandgap transition by comparing the experimentally obtained emission peak position and the calculated emission wavelengths for different strain values (Fig. S2, supplementary material). The redshift of the PL signal from the unreleased to the released nanobeam is attributed to the decreased direct bandgap energy in the released nanobeam with lower compressive strain.[16] The released nanobeam reveals an onset of lasing with a pump power density of 14.5 kW cm$^{-2}$ (green curve in Fig. 2(b)), whereas the unreleased nanobeam only shows a broad spontaneous emission spectrum even at a higher pump power density of 16.3 kW cm$^{-2}$ (black curve in Fig. 2(a)). In Figure 2(c), we show the light-in-light-out (L-L) curves of both unreleased and released nanobeams to compare the lasing characteristics. The double-logarithmic plots in the inset show typical S-shaped curves for both devices, implying clear lasing behaviors. The threshold density of 38.4 kW cm$^{-2}$ and 18.2 kW cm$^{-2}$ are measured for the unreleased and released nanobeams, respectively. The reduced threshold in the released nanobeam can be attributed to the improved directness owing to the absence of compressive strain. The increased directness leads to a larger electron population in the direct Γ conduction valley, resulting in a higher optical gain.[43] Figure 2(d) presents the full-width at half-maximum (FWHM) of both nanobeams as a function of pump power density. Both devices show a reduction in the FWHM from ~2.5 nm to ~0.8 nm, which provides further evidence of lasing.

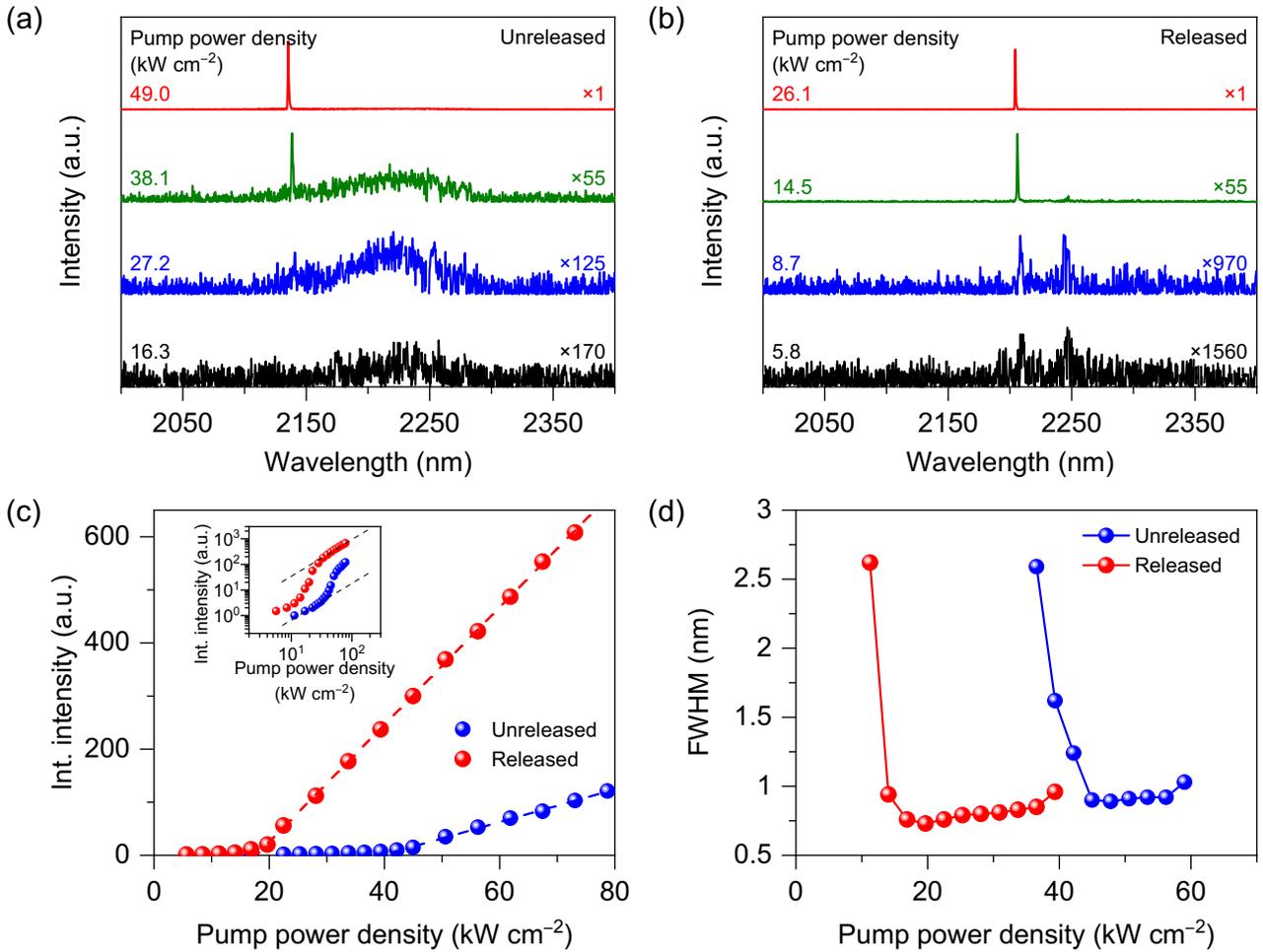

FIG. 2. PL spectra measured at 4 K for the (a) unreleased and (b) released nanobeams under various pumping powers. (c) L-L curves for the unreleased and released nanobeams. Corresponding double-logarithmic plot is shown in the inset. (d) FWHM vs. pump power density for the unreleased and released nanobeams at 4 K.

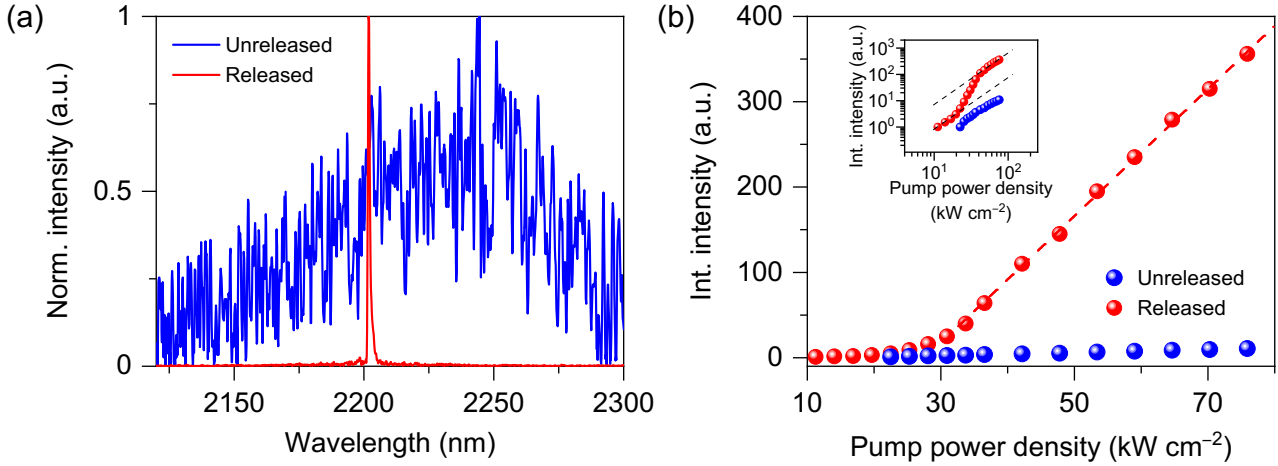

FIG. 3. (a) Normalized PL spectra of the unreleased and released nanobeams at 70 K. (b) Power-dependent PL spectra of the unreleased and released nanobeams at 70 K. The double-logarithmic plot for the same dataset is shown in the inset.

Figure 3(a) shows normalized PL spectra of the unreleased (blue curve) and released (red curve) nanobeams. The measurement temperature and the pump power density were 70 K and 50.6 kW cm$^{-2}$, respectively. While the lasing action persists in the released nanobeam, only a broad spontaneous emission is observed for the unreleased nanobeam. The contrasting behaviors of the two nanobeams can also be observed from the L-L curves, as shown in Fig. 3(b). A clear lasing behavior of the released nanobeam can be evidenced by the S-shaped curve in the inset of Fig. 3(b). In contrast, the absence of lasing in the unreleased nanobeam is revealed by the linearity of the L-L curve for all pumping powers. The higher operating temperature of the released nanobeam is attributed to the improved directness of the released nanobeam, which is explained in more detail in the next section.

The lasing characteristics of the released nanobeam were further investigated with temperature-dependent PL measurements, as shown in Fig. 4(a). The PL spectra were taken with a fixed pump power density of 33.7 kW cm$^{-2}$ at various temperatures between 4 and 110 K. The intensity of the lasing mode is decreased at higher temperatures, and only a broad spontaneous emission is observed at 110 K (black curve). Figure 4(b) presents L-L curves obtained for the same temperature range. The inset is plotted on double-logarithmic scales for the same dataset. Typical non-linear S-shaped curves are acquired at temperatures up to 90 K, whereas the integrated intensity grows linearly with increasing pump power at 110 K. We conducted theoretical modelling using k·p method for GeSn with a Sn content of 10.6 at% to further study the quenching of lasing at 110 K. Figure 4(c) presents optical net gains as a function of temperature with injected carrier densities of $1.0 \times 10^{17}$ cm$^{-3}$, $5.0 \times 10^{17}$ cm$^{-3}$, and $1.0 \times 10^{18}$ cm$^{-3}$, which correspond to pump power densities of 3.8 kW cm$^{-2}$, 18.7 kW cm$^{-2}$, and 37.4 kW cm$^{-2}$, respectively. The calculations show that the net gain increases at higher pump powers, but decreases with increasing device temperature. Quenching of lasing occurred below 180 K although the positive net gain remains up to ~180 K at an injected carrier density of $1.0 \times 10^{18}$ cm$^{-3}$. The quenched lasing behaviour at 110 K can be attributed to the reduced carrier density in the Γ conduction valley at elevated temperatures and activation of the nonradiative process, resulting in a decreased optical net gain.[15,29,44]

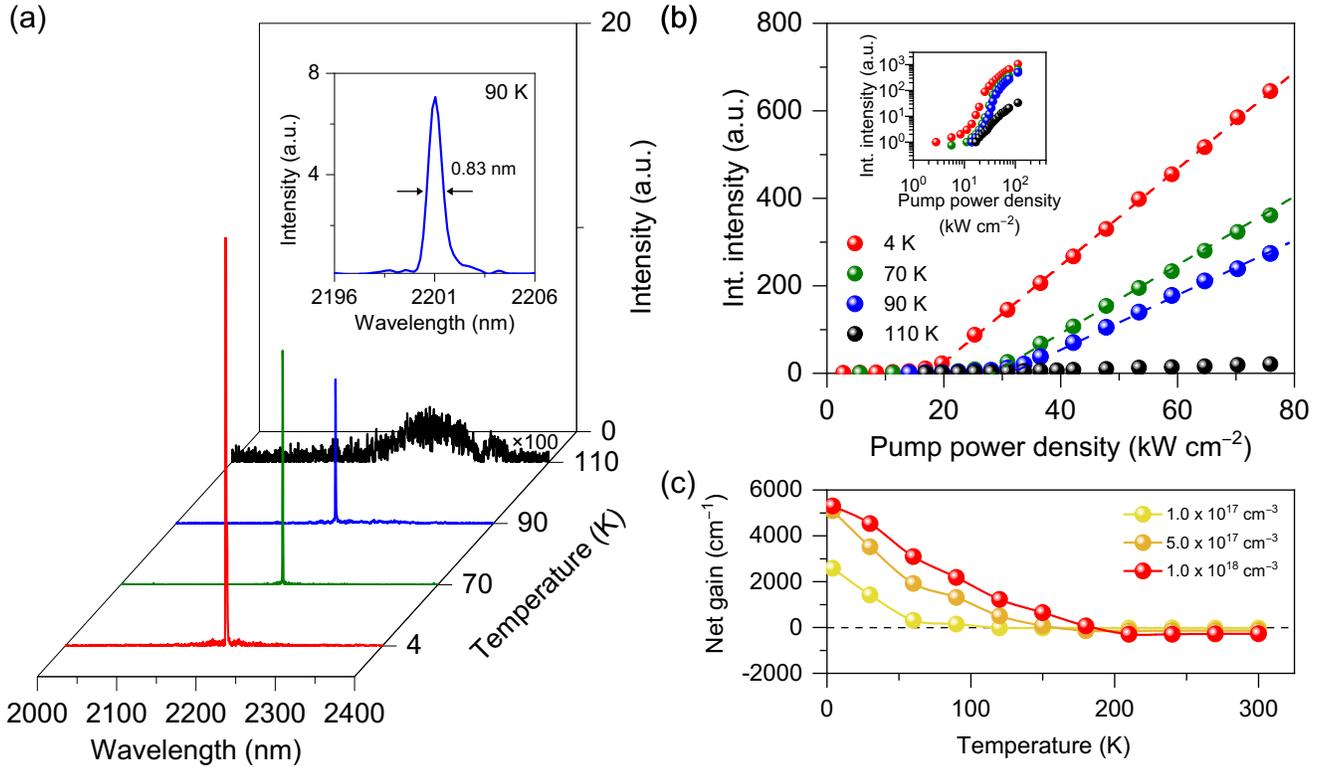

FIG 4. (a) Temperature-dependent PL spectra of the released nanobeam under a pump power density of 33.7 kW cm$^{-2}$. (b) L-L curves for the released nanobeam measured at various temperatures. Corresponding double-logarithmic plot is shown in the inset. (c) Optical net gain for various temperatures at injected carrier densities of $1.0 \times 10^{17}$ cm$^{-3}$ (yellow), $5.0 \times 10^{17}$ cm$^{-3}$ (orange), and $1.0 \times 10^{18}$ cm$^{-3}$ (red).

In summary, we have demonstrated lasing characteristics of 1D PC nanobeams fabricated on a GeSnOI substrate with a very small device footprint of 7 μm$^2$. An improved directness enabled by releasing the limiting compressive strain allowed achieving a lower threshold density of 18.2 kW cm$^{-2}$ at 4K compared to the unreleased nanobeam laser exhibiting a threshold density of 38.4 kW cm$^{-2}$ at the same temperature. The released nanobeam showed clear lasing behavior up to 90 K, whereas the unreleased nanobeam stopped lasing at a temperature lower than 70 K. The device configuration we demonstrate in this work may be particularly useful in achieving electrically driven GeSn lasers operating at a low power consumption because the small active area of the 1D PC nanobeam lasers requires a very low current injection to make the active gain medium transparent.[38] An electrical injection can be easily achieved in our nanobeam structure through the lateral p-i-n junction design as demonstrated by Shambat et al.[45] We believe that our demonstration provides opportunities to realize practical laser sources with dense integration and low power consumption for CMOS-compatible PICs.

## SUPPLEMENTARY MATERIAL

See supplementary material for the pump-power-dependent Raman spectroscopy; and simulated directness and emission wavelength.

## ACKNOWLEDGMENTS

The research of the project was in part supported by Ministry of Education, Singapore, under grant AcRF TIER 1 2019-T1-002-050 (RG 148/19 (S)). The research of the project was also supported by Ministry of Education, Singapore, under grant AcRF TIER 2 (MOE2018-T2-2-011 (S)). This work is also supported by National Research Foundation of Singapore through the Competitive Research Program (NRF-CRP19-2017-01). This work is also supported by National Research Foundation of Singapore through the NRF-ANR Joint Grant (NRF2018-NRF-ANR009 TIGER). This work is also supported by the iGrant of Singapore A*STAR AME IRG (A2083c0053). OM acknowledges support from NSERC Canada (Discovery, SPG, and CRD Grants), Canada Research Chairs, Canada Foundation for Innovation, Mitacs, PRIMA Québec, and Defence Canada (Innovation for Defence Excellence and Security, IDEaS).

## DATA AVAILABILITY

The data that support the findings of this study are available from the corresponding author upon reasonable request.

# Supporting information:

## 1D photonic crystal direct bandgap GeSn-on-insulator laser

Hyo-Jun Joo,[1,a)] Youngmin Kim,[1,a)] Daniel Burt,[1] Yongduck Jung,[1] Lin Zhang,[1] Melvina Chen,[1] Samuel Jior Parluhutan,[1] Dong-Ho Kang,[1] Chulwon Lee,[2] Simone Assali,[3] Zoran Ikonic,[4] Oussama Moutanabbir,[3] Yong-Hoon Cho,[2] Chuan Seng Tan,[1] and Donguk Nam[1,b)]

[1] School of Electrical and Electronic Engineering, Nanyang Technological University, 50 Nanyang Avenue, Singapore 639798, Singapore

[2] Department of Physics and KI for the NanoCentury, Korea Advanced Institute of Science and Technology (KAIST), Daejeon 34141, Republic of Korea

[3] Department of Engineering Physics, École Polytechnique de Montréal, C. P. 6079, Succ. Centre-Ville, Montréal, Québec H3C 3A7, Canada

[4] School of Electronic and Electrical Engineering, University of Leeds, Leeds LS2 9JT, UK

## Contents

# 1. Pump-power-dependent Raman spectroscopy

Figure S1(a) shows the shift of the relative Raman peak position as a function of the pump power for both the unreleased and released nanobeams. In both devices, it was confirmed that the Raman shift occurred above the pump power of 200 µW. The Raman spectra in Fig. 1(c) was obtained at a pump power of 160 µW to avoid heating effect on the devices. Figure S1(b) presents the FWHM of Raman spectra as a function of pump power for both nanobeams, showing no significant broadening of the FWHM at the pump power of 160 µW.

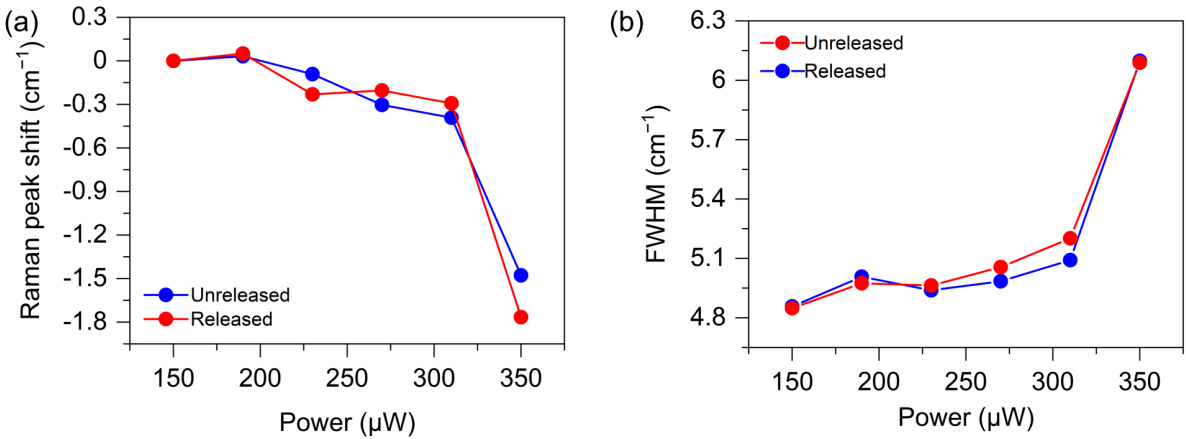

Figure S1. Pump-power-dependent (a) Raman peak shifts and (b) FWHM for the unreleased and released nanobeams.

## 2. Simulated directness and emission wavelength

Figure S2(a) presents the energy difference between the L and Γ conduction band-edge minima (i.e., the directness) as a function of the biaxial strain. The calculated direct bandgap energy of the unreleased nanobeam which has an intrinsic compressive strain of −0.23% is ~0.549 eV. For the released nanobeam which has a fully relaxed strain, the direct bandgap energy is ~0.541 eV. Figure S2(b) shows the calculated direct bandgap emission wavelength. The unreleased and released nanobeams show direct bandgap emission wavelengths of ~2260 nm and ~2290 nm, respectively. The measured direct bandgap emission wavelengths for the unreleased (~2240 nm) and released (~2270 nm) nanobeams are in good agreement with the corresponding simulated values.

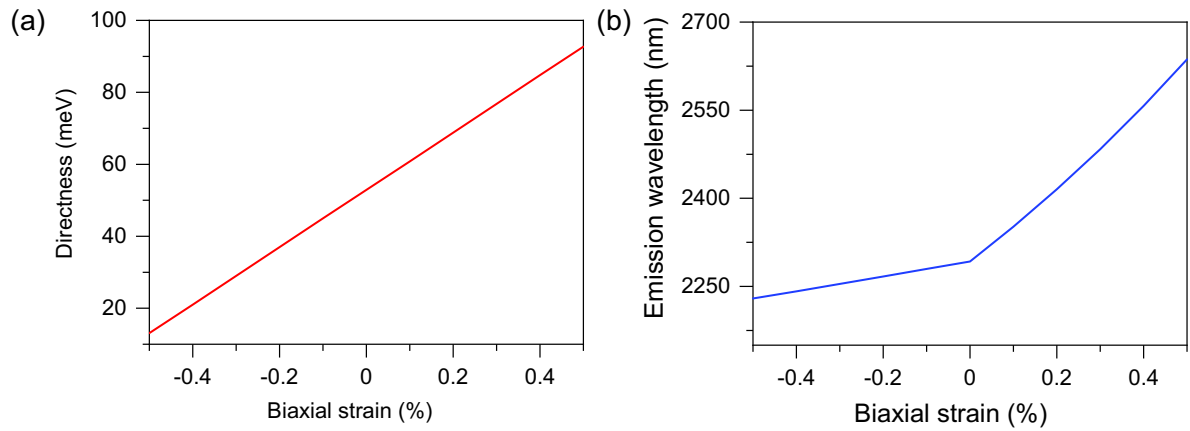

Figure S2. (a) Simulated directness and (b) emission wavelength of GeSn with 10.6 at% as a function of biaxial strain.